\documentclass[%
 reprint,
superscriptaddress,
 amsmath,amssymb,
  aps,
]{revtex4-1}

\usepackage{graphicx}% Include figure files
\usepackage{dcolumn}% Align table columns on decimal point
\usepackage{bm}% bold math
\usepackage{hyperref}% add hypertext capabilities
\begin{document}

%\preprint{APS/123-QED}

\title{
White-light-seeded, CEP-stable, 4.5-W, 4-micron KTA
parametric amplifier driven by a 1.4-ps Yb:YAG thin disk laser
}


\author{Tsuneto Kanai}
 \email{goldwell74@gmail.com}
\affiliation{Department of Physics \& Center for Attosecond Science and Technology, 
Pohang University of Science and Technology, Pohang 37673, South Korea}
\affiliation{Max Planck Center for Attosecond Science, Max Planck POSTECH/Korea Research Initiative, Pohang 37673,
South Korea}

\author{Yeon Lee}
\affiliation{Department of Physics \& Center for Attosecond Science and Technology, 
Pohang University of Science and Technology, Pohang 37673, South Korea}
\affiliation{Max Planck Center for Attosecond Science, Max Planck POSTECH/Korea Research Initiative, Pohang 37673,
South Korea}

\author{Meenkyo Seo}
\affiliation{Department of Physics \& Center for Attosecond Science and Technology, 
Pohang University of Science and Technology, Pohang 37673, South Korea}
\affiliation{Max Planck Center for Attosecond Science, Max Planck POSTECH/Korea Research Initiative, Pohang 37673,
South Korea}

\author{Dong Eon Kim}
\affiliation{Department of Physics \& Center for Attosecond Science and Technology, 
Pohang University of Science and Technology, Pohang 37673, South Korea}
\affiliation{Max Planck Center for Attosecond Science, Max Planck POSTECH/Korea Research Initiative, Pohang 37673,
South Korea}


\begin{abstract}
We demonstrate a robust, carrier envelope phase (CEP)-stable, potassium titanyl arsenate (KTA)-based 
optical parametric amplifier (OPA) delivering 6-cycle (79 fs), 3.8-$\mu$m pulses  
at a 100-kHz repetition rate with an average power of 4.5 W. 
The pivotal achievement is stable generation of supercontinuum (SC) seed pulses in a YAG crystal
with a rather long pulse of 1.4 ps; to our knowledge, this is the longest duration for SC generation (SCG).
This technology offers a robust and simplified OPA architecture 
with characteristics of passively-stabilized CEP,
simplified dispersion management with bulk materials, 
wavelength tunability of the output pulses from 1.3-4.5 $\mu$m, 
and the future power scaling up to kW-class based on Yb:YAG thin disk amplifiers. 
The total output power of 17 W (signal plus idler)  is achieved 
and the capability of this high photon flux aspect is successively demonstrated by its application 
to high harmonic generation (HHG) in ZnSe crystals, 
with which faint yet novel signals above their bandgap are clearly observed.
%\begin{description}
%\end{description}
\end{abstract}

\maketitle

%\tableofcontents

  Driven by a guiding principle, $\lambda^2$ energy scaling law
 of strong field phenomena \cite{popmintchev2012bright}, where $\lambda$ is the wavelength of the driving pulses,
development of high power femtosecond lasers in 3-10 $\mu$m region 
\cite{Andriukaitis2011,Kanai:17,vonGrafenstein:17,Fan:16,Kanai:18,Archipovaite:17,
Elu:17,Thire:17,Baudisch:14,Rigaud:16,Seideleaaq1526,Mero:15} attract much attention 
as next generation sources in ultrafast science and its related areas.
Ones of the most successful lasers of this kind are relatively-low-power but high-pulse-energy KTA OPAs 
at the 4-$\mu$m band. 
The 3.9 $\mu$m KTA OPA with 0.16 W and 8 mJ Ref.~\cite{Andriukaitis2011} was used to
realize coherent 1.6 keV hard X-ray generation via HHG \cite{popmintchev2012bright}, 
laser-plasma hard X-ray emission \cite{weisshaupt2014high}, 
and mid-IR (MIR) filamentation \cite{mitrofanov2015mid}.
Another aspect of this scaling law, whose principle is in the criteria 
of non-perturbative phenomena, Keldysh parameter 
$\gamma :=\sqrt{I_\mathrm{p}/2U_\mathrm{p}}\propto \lambda^{-1} I^{-1/2} $ 
($I_\mathrm{p}$: ionization potential, $U_\mathrm{p}$: ponderomotive potential, 
$I$: intensity of the laser pulses),
is the realization of non-perturbative phenomena in fragile and functional materials 
such as semiconductors \cite{ghimire2011observation,Chin2001,Kanai:17}, bio-molecules \cite{Pupeza2015}, and so on;
with MIR pulses, one can realize $\gamma \lesssim 1$ while keeping the laser intensity $I$ below their damage threshold.

While the latter applications started to offer novel methodologies for material science and bio-physics, 
the required laser performance is quite different, 
i.e.~, both high-pulse-energy ($\gtrsim 10$ $\mu$J) and high-power ($\gtrsim $ 1W) are crucial.
However thermal effects by single- and/or multi-photon absorption in the typical oxide-type nonlinear crystals become serious for developing these lasers  \cite{Baudisch:14}.
Recently some lasers of this class started to be reported at 3-$\mu$m-band with niobate-based OPAs \cite{Thire:17,Rigaud:16,Mero:15,Elu:17},
however, elongating their wavelength to the 4-$\mu$m band has not been achieved to the best of our knowledge.

%%%%%%%%%%%%%%%%%%%%%%%%%%%%%%%%%%%%%%%%%%
\begin{figure*}[t]
 \centering
 \includegraphics[width=1\linewidth]{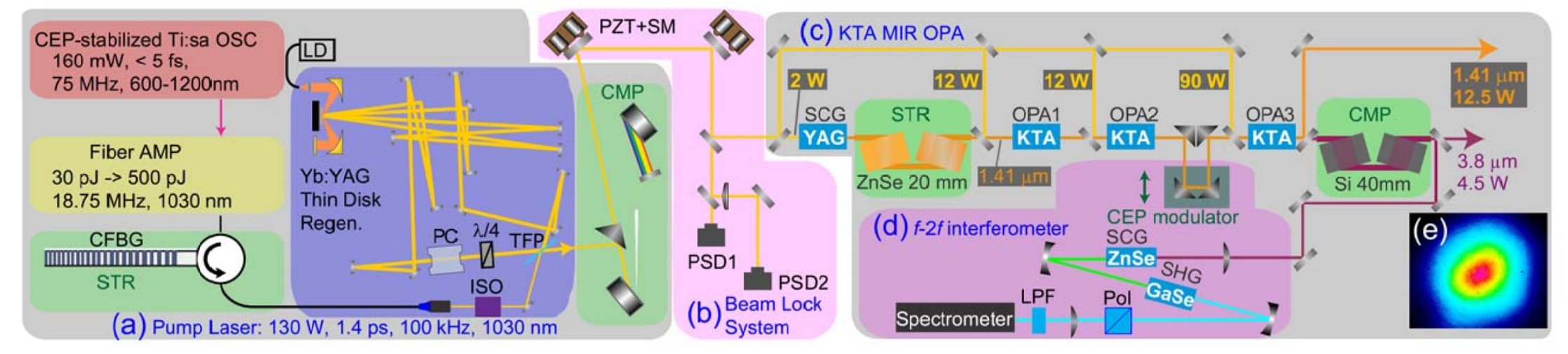}
 \caption{%
 \small{%
The optical layout composed of a Yb:YAG thin disk amplifier seeded by a CEP-stabilized Ti:sa oscillator (a), 
a beam locking system (b), 
a KTA OPA (c), an $f$-$2f$ interferometer with a CEP modulator (d). 
(e) A typical beam pattern of the 3.8 $\mu$m pulses observed 
with a MIR bolometer array camera (WinCamD-IR-BB, DataRay Inc.).
Ti:sa: Titanium-sapphire; OSC: oscillator; AMP: amplifier; CMP: compressor; 
CFBG: chirpted fiber Bragg grating; ISO: isolator, PC: Pockels Cell; TFP: thin film polarizer; WP: wave plate; 
PZT: piezoelectric actuator; PSD: position-sensitive detector.
 }
 }
\label{fig1}
\end{figure*}
%%%%%%%%%%%%%%%%%%%%%%%%%%%%%%%%%%%%%%%%%%
%%%%%%%%%%%%%%%%%%%%%%%%%%%%%%%%%%%%%%%%%%
\begin{figure}[!t]
 \includegraphics[width=\linewidth]{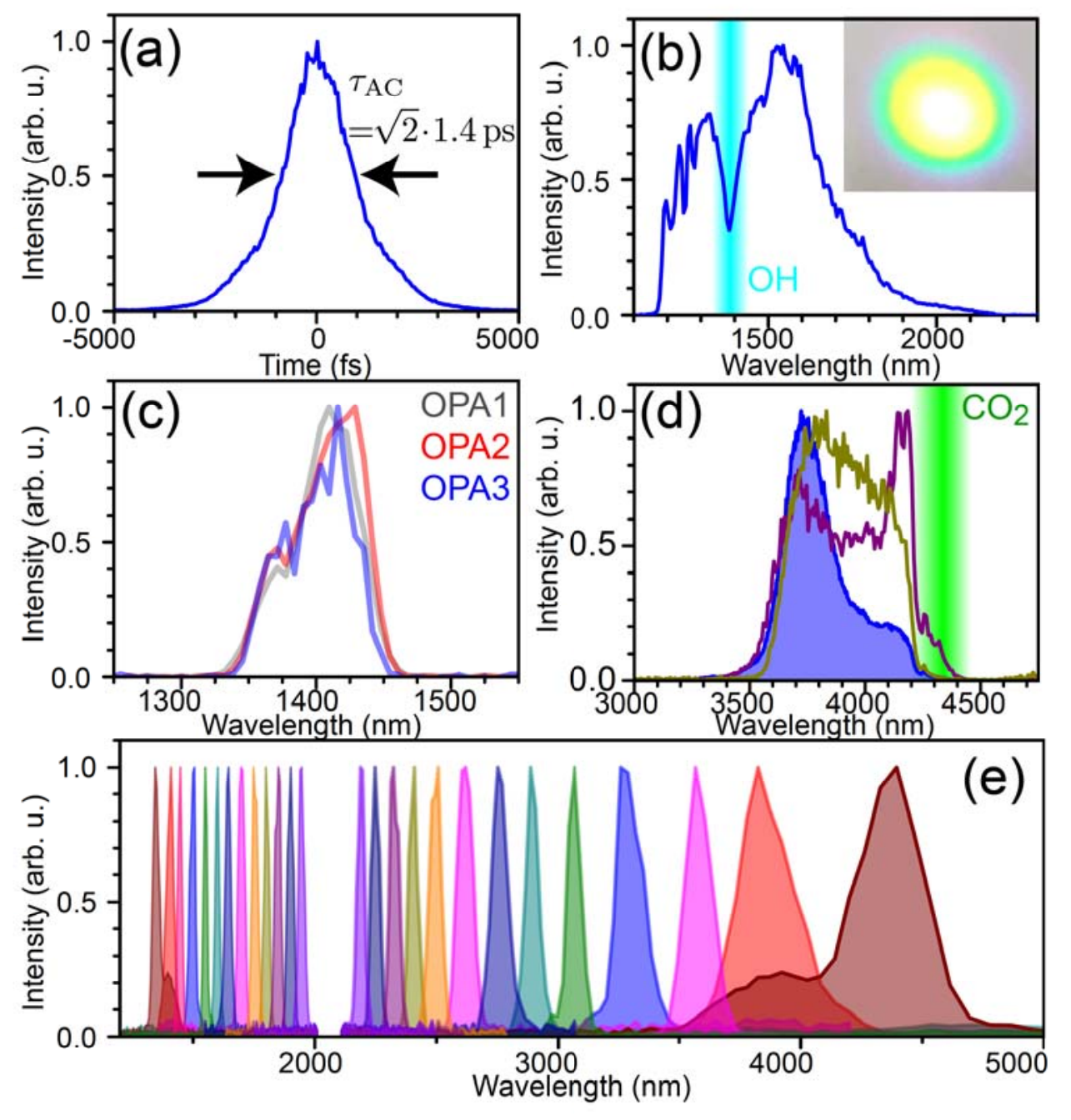}
 \caption{%
 \small{%
(a) Autocorrelation trace of the thin disk laser outputs. 
Autocorrelation width $\tau_\mathrm{AC}$ was 1.95 ps, which corresponds to the pulse duration of 1.4 ps 
under the assumption of Gaussian pulses.  
The side wings
originate from uncompressible high order dispersion from the CFBG. 
(b) A typical spectrum of the SC generated in a 10-mm-thick YAG crystal.  
Here the modulation around 1.2 $\mu$m and the dip at 1.4 $\mu$m are from a 1 $\mu$m filter and -OH absorption in a fiber, respectively.
A beam image taken with a visible camera is also shown in the inset. 
(c) Typical spectra of the signals of the KTA OPA at each stage.  
(d) A typical idler spectrum of KTA OPA (blue curve) measured with acousto-optic-modulator-based MIR spectrometer (MOZZA, Fastlite). 
The purple and dark yellow curves with broader spectra were obtained at different pumping timing 
with the same phase matching. 
(e) Wavelength tunability of the laser system.  The signal covers from 1.3 $\mu$m to 2.0 $\mu$m and the idler
covers from 2.1 $\mu$m to 4.5 $\mu$m.
 }
 }
\label{fig2}
\end{figure}
%%%%%%%%%%%%%%%%%%%%%%%%%%%%%%%%%%%%%%%%%%

Another challenge of MIR OPA laser development is to clarify whether we can adopt the SC-seeding scheme, which simplifies OPA architecture dramatically using available pump lasers such as 
Yb:fiber lasers \cite{Archipovaite:17}, 
Yb:YAG disk lasers \cite{Fattahi:14,Giesen1994,Nubbemeyer:17,Prinz:15,Pupeza2015}, 
Nd-based lasers \cite{Andriukaitis2011,Elu:17}, 
Ho-based lasers \cite{Kanai:17,vonGrafenstein:17,Kanai:18}, and so on.
In fact, similar to the history of Ti:sa-driven OPAs \cite{shirakawa1998noncollinearly}, 
some of Yb-based femtosecond lasers adopted this scheme and realized simple and robust OPAs without broadband oscillators nor pump-seed-synchronization setups \cite{Kanai:17,Fan:16,Archipovaite:17,Rigaud:16}.
The exact conditions for the adoption of SC scheme, however, have not been clear 
especially for the most developing Yb:YAG thin disk amplifiers \cite{Fattahi:14,Giesen1994,Nubbemeyer:17,Prinz:15}, 
since their typical pulse duration is in the picosecond region, where SCG had been long considered to be difficult because 
of the injurious competing avalanche ionization process in bulk \cite{Calendron:15}.
Recently, Thir\'e \textit{et al}.~demonstrated an OPA at the 3-$\mu$m band based on MgO-doped periodically poled lithium
niobate crystal (MgO:PPLN) with SC being generated by 1.1 ps pulses \cite{Thire:17}.
While they report the output power of 4 W at 3.1 $\mu$m, the output power 
 is dramatically reduced to 0.5 W in the 4-$\mu$m band, which indicates the difficulty 
in high power laser development in this band.

In this paper, we demonstrate an SC-seeded, 4.5 W, 100 kHz, 3.8 $\mu$m KTA OPA 
driven by a 1.4 ps Yb:YAG thin disk amplifier.  
The pivotal achievement is stable generation of SC seed pulses in a YAG crystal
with 1.4 ps pulses, whose duration is the longest for SCG 
and typical for Yb-YAG thin-disk amplifiers \cite{Nubbemeyer:17}.
The current development offers (1) simple and robust OPA architecture, 
(2) passively stabilized CEP through difference frequency generation (DFG) \cite{Baltuska2002},
(3) simplified dispersion management with bulk materials, 
(4) wavelength tunability of the output pulses in 1.3-4.5 $\mu$m, 
and (5) the future power scaling with kW-class Yb:YAG thin disk amplifiers \cite{Nubbemeyer:17}. 
The total output power of 17 W (signal plus idler) is achieved 
and the capability of this laser was successively proved by a demonstration experiment on HHG in ZnSe crystals, where faint yet novel signals above its bandgap are clearly observed.

Figure \ref{fig1}(a-d) shows the optical layout composed of a Yb:YAG thin disk amplifier 
(modified TruMicro 5070, Trumpf Scientific Lasers GmbH + Co.~KG) 
seeded by a CEP-stabilized Ti:sa oscillator (a), 
a 4D-beam locking system (Aligna, TEM Messtechnik GmbH) (b), 
a KTA OPA (c), an $f$-$2f$ interferometer with a CEP modulator based on a stepper motor (SM)-driven delay stage (d).   
The thin-disk laser delivers 130W at 100 kHz on target and its pulse duration was measured
with a BBO-based second harmonic generation (SHG) autocorrelator to be 1.4 ps [Fig.~\ref{fig2}(a)].  
While the thin disk laser is capable of 200 W operation, we saved the 70 W for the 
future upgrading.
After reducing their beam pointing fluctuation [Fig.~\ref{fig1}(b)], mainly from the thermal drift  
and air turbulence during beam delivery for $\approx 4$ meters, 
a portion of the output pulses with 2 W power was used to generated stable SC up to $\approx 2.2$  $\mu$m in a 10-mm-thick undoped YAG crystal [Fig.~\ref{fig2}(b)]. 
Here we carefully optimized SCG with a relatively loose focusing geometry and a hard edge aperture (numerical aperture $\approx 0.0025$)
and confirmed its stability with longer-duration pump pulses up to $\approx 1.6$ ps.

\begin{figure}[!t]
 \centering\includegraphics[width=\linewidth]{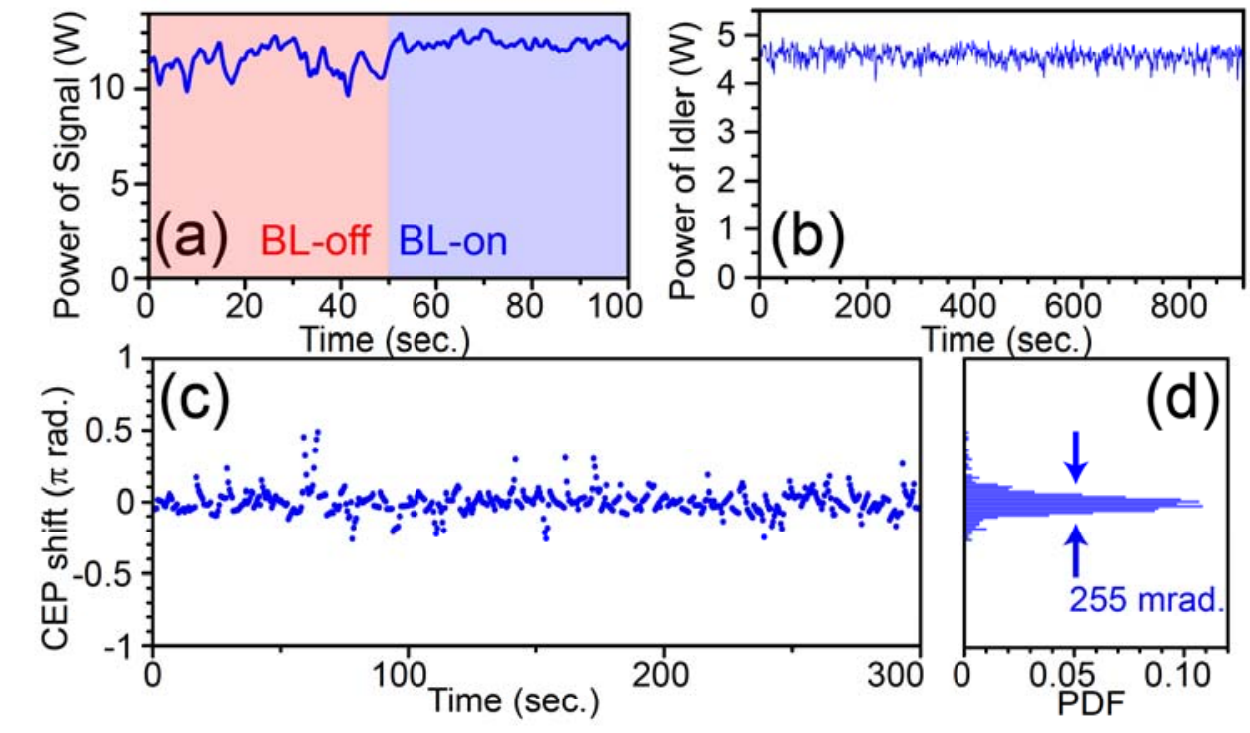}
 \caption{%
 \small{%
(a) The power stability of signal pulses with (SD: 2.0 \%) and without (SD: 6.0 \%) beam locking (BL). 
(b) The power stability of idler pulses with beam locking (SD: 2.0 \%) . 
 (c) Passively stabilized CEP shifts as a function of time. (d) CEP shift histogram, 
which shows its SD 255 mrad. Here PDF denotes probability density function.}
 }
\label{fig3}
\end{figure}

The full temporal coherence of the SC was confirmed by its amplification by the three-stage KTA OPA (OPA1,2,3) with
ultrabroad wavelength tunability presented in Fig.~\ref{fig2}(e) and the typical spectra used in the 4-$\mu$m-band operation for each OPA stage 
are depicted in Fig.~\ref{fig2}(c,d).
Here the 1.41-$\mu$m component of SC pulses was first temporally stretched with a pair of ZnSe Brewster plates 
 to match its pulse duration with that of the pump, 
and amplified in OPA1 and OPA2,
where 5-mm-thick and 10-mm-thick AR-coated Type II KTAs ($\theta=40.5^\circ$, $\phi=0^\circ$) were used as nonlinear crystals.
Here the pumping power for both stages were 12 W and output power of the signal at 1.41-$\mu$m from OPA2 was typically 500 mW.
In OPA3 with a 5-mm-thick Type II KTA with the same phase matching angles, this 1.41-$\mu$m pulses were injected as seeds, 
resulting in the generation of 12.5 W, 1.41 $\mu$m signal pulses and 4.5 W, 3.8-$\mu$m idler pulses with 2 \% SD [Fig.~\ref{fig3}(a,b)], which was from the fluctuation (2 \% SD) of the pump power.
The pumping power for OPA3 was set to be 90 W; hence the total MIR conversion efficiency was 13 \% at total output power 17 W, which is similar
 to those of other KTA OPAs, e.g., 14\% at total output power of 0.7 W in Ref.~\cite{Andriukaitis2011} 
and 12.5 \%  at total output power of 0.75 W in Ref.~\cite{Fan:16}.
Here to keep the beam profile undistorted thermally [Fig.~\ref{fig1}(e)], 
the pumping peak intensity for all of the stages were set to $57$ GW/cm$^2$, similar value reported in Ref.~\cite{Baudisch:14} and was 20$\%$ below the damage threshold of the AR coating.
The CEP of the idler was passively stabilized through DFG at OPA3 between the pump 
and the 1.41-$\mu$m seed pulses from the same origin 
and its spectrum is temporally selected to shorter wavelength to avoid the dispersion and absorption 
by KTA itself and CO$_2$ in the air around 4255 nm [Fig.~\ref{fig2}(d)]. 
%
%%%%%%%%%%%%%%%%%%%%%%%%%%%%%%
%%%%%%%%%%%%%%%%%%%%%%%%%%%%%

%
%

The stability of the CEP of the idler was measured with an inline $f$-$2f$ interferometer [Fig.~\ref{fig1}(d)].
The 3.8 $\mu$m pulses drove SCG in a 3-mm-thick ZnSe crystal, which spanned from 2 to 5 $\mu$m, 
and its 4.8-$\mu$m component 
was frequency doubled in a 1-mm-thick Type I GaSe crystal ($\theta=10.8^\circ$) and interfered 
with the shorter component of the SC around 2.4 $\mu$m. 
The amplitude ratio of fundamental pulse and SH pulses was adjusted by rotating a polarizer 
and the third harmonic (TH) of the fundamental beam generated in the ZnSe crystal was blocked by a long-pass filter.  
 Figure \ref{fig3}(c) shows measured typical CEP shifts as a function of time and its standard deviation (SD) was measured to be 255 mrad.  Here we also confirmed the controllability of the CEP by moving the delay stage between seed and pump 
for OPA3 [Fig.~\ref{fig1}(d)] for future applications.

%%%%%%%%%%%%%%%%%%%%%%%%%%%%%%%%%%%%%%%%%%
\begin{figure}[!t]
 \includegraphics[width=\linewidth]{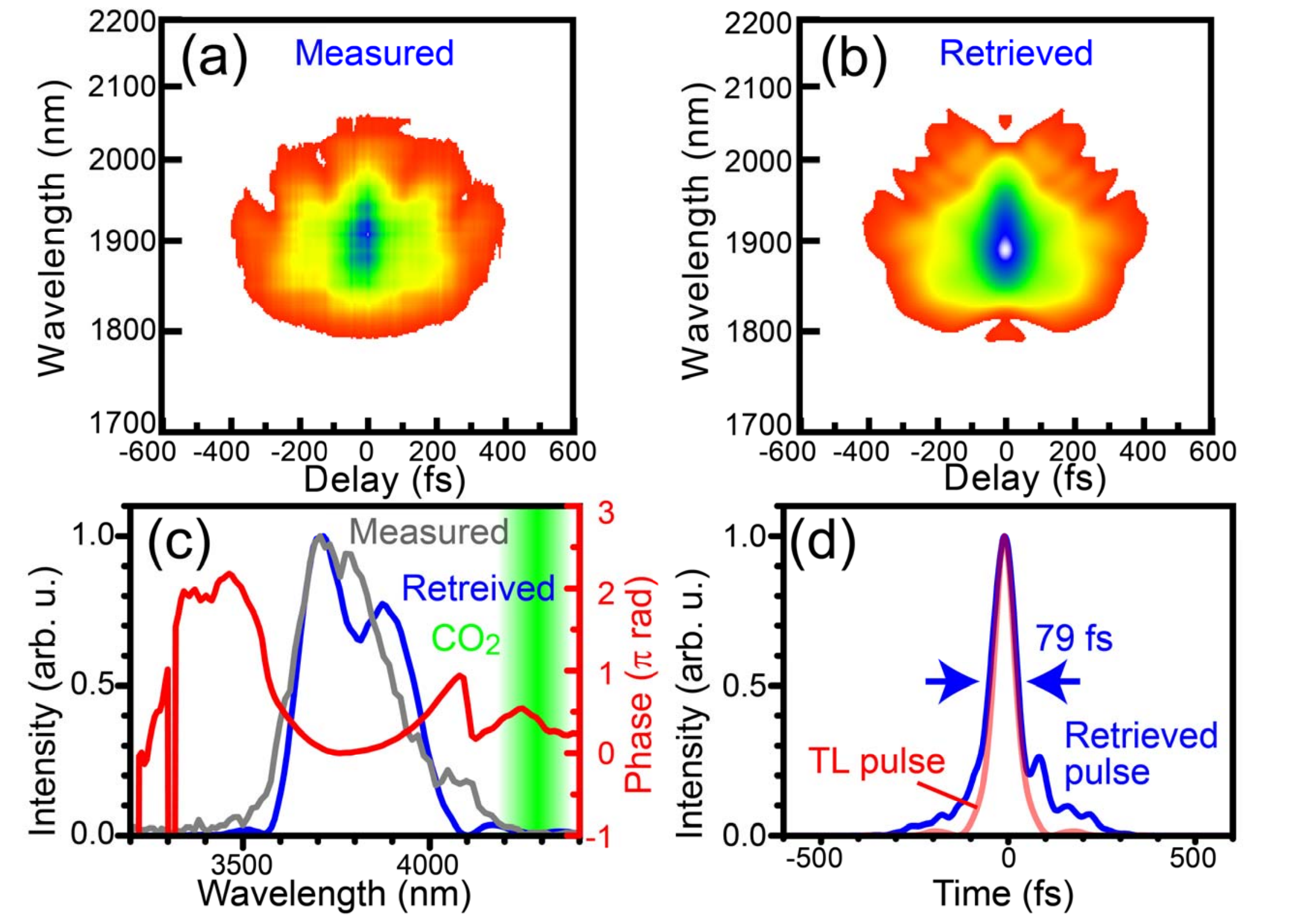}
 \caption{%
 \small{%
SHG-FROG characterization of the 79-fs, 3.8-$\mu$m pulses from the KTA OPA.  
Measured (a) and reconstructed (b) SHG-FROG traces.
 (c) Retrieved spectrum (blue curve), retrieved spectral phase (red curve), 
and the spectrum measured with the MIR spectrometer (gray curve).
 (d) Retrieved temporal pulse profile (blue curve, 79 fs) 
with the Transform Limited (TL) pulse (red curve, 70 fs). 
 }
 }
\label{fig4}
\end{figure}
%%%%%%%%%%%%%%%%%%%%%%%%%%%%%%%%%%%%%%%%%%
The last advantage of the SC-seeding scheme is the availability to use bulk stretcher(s) and/or compressor(s) 
for its dispersion management, which minimizes power losses and CEP noises compared to those 
with grating-based or prism-based dispersion control. 
Here we adopted ZnSe-Brewster plates [20 mm, GDD (group delay dispersion): $+8 \times 10^3$ fs$^2$] for stretching 1.41 $\mu$m seed pulses and Si-Brewster plates  \cite{Kanai:17} (40 mm, GDD: $+1.6 \times 10^4$ fs$^2$)
for compressing amplified 3.8 $\mu$m from OPA3.
Intrinsic GDD in SCG was measured to be $+8 \times 10^3$ fs$^2$ and by considering the phase conjugation between the signal and idler waves at OPA3 [Fig.~\ref{fig1}(c)],
the GDD of the 3.8 $\mu$m is expected be negligible.
Here both the stretcher and the compressor have $>90$\% throughputs, suppressed 
B-integrals through lateral expansion in the bulk compared to the normal incidence case, 
and small chirps  geometrically-induced between the plates.

This simple dispersion control scheme was proved by 
a MIR SHG-FROG (Frequency-resolved optical gating) setup using a 30-$\mu$m-thick Type I GaSe ($\theta =12.3^\circ$) SHG crystal 
and an InGaAs-based NIR spectrometer (NIRQuest256, Ocean Optics Inc.). 
Figure \ref{fig4} summarizes the pulse compression results 
and the measured compressed pulse FWHM duration was 79 fs ($\approx 6$ optical cycles) 
whereas the amplified spectrum supports a 70-fs duration. 
Here the TL duration for the broadest spectrum [Fig.~\ref{fig2}(d), purple curve] was 49 fs,
which was elongated to $> 100$ fs due to the above mentioned dispersion from KTA and CO$_2$ 
and its further compression is under investigation. 

\begin{figure}[!t]
 \centering
 \includegraphics[width=\linewidth]{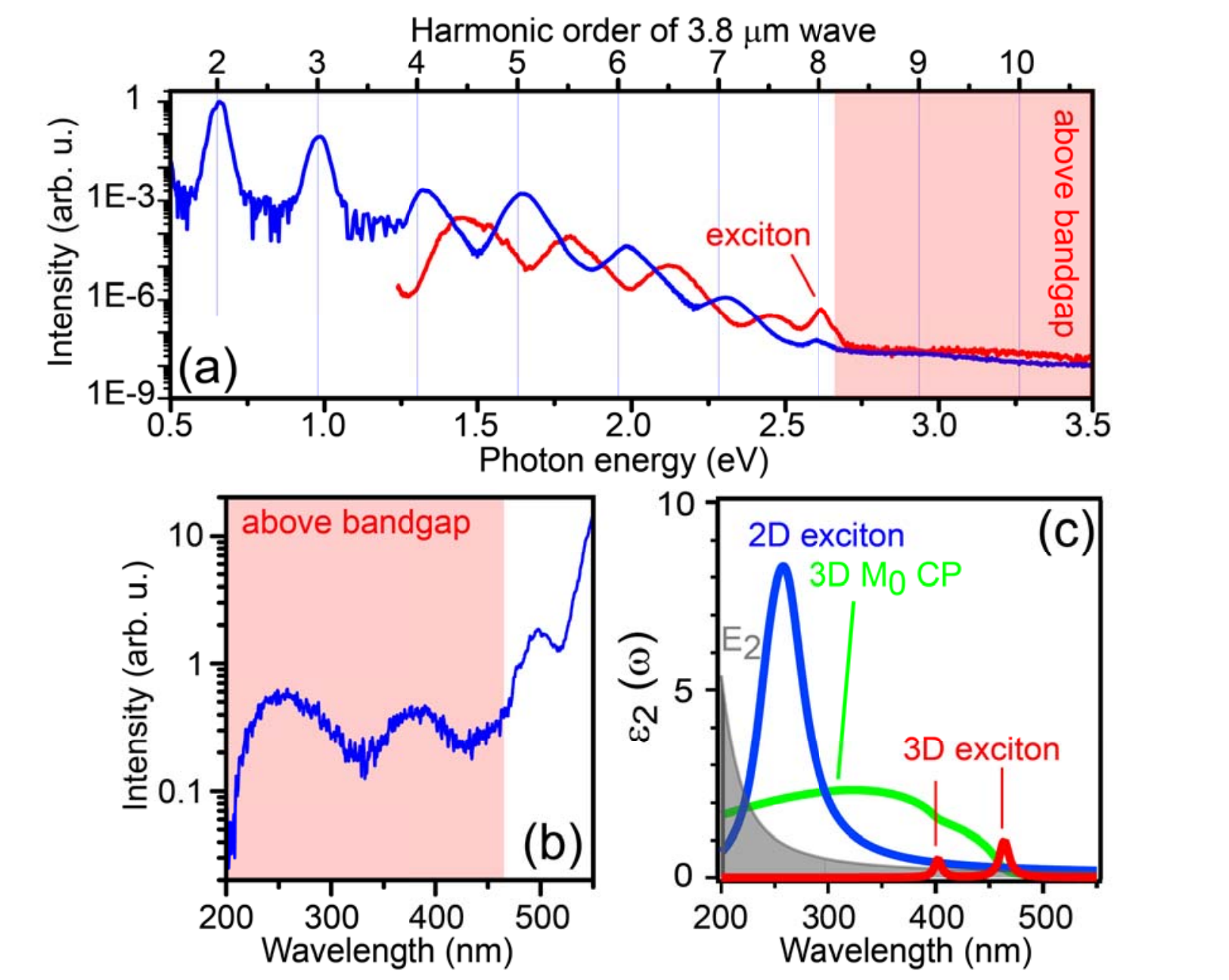}
 \caption{%
 \small{%
 (a) A typical spectrum of high harmonics driven by 3.8 $\mu$m (blue curve) 
and 3.5 $\mu$m (red curve) pulses in a 3-mm-long ZnSe polycrystal 
measured with a CCD (200--1000 nm) and an InGaAs (1000--2500 nm) spectrometers. 
(b) Observed spectra above bandgap with the optimized two peaks at $\approx 250$ nm and $\approx 400$ nm.
(c) Theoretically calculated individual contributions to $\epsilon_2(\omega)$
 of ZnSe from the 3D $M_0$ CP (critical point) and the 3D-exciton at $E_0$ and $E_0+\Delta_0$, 
the 2D-exciton at $E_1$ and $E_1+\Delta_1$, 
and the $E_2$-gap \cite{PhysRevB.14.556,Adachi1991}. 
 }
 }
\label{fig5}
\end{figure}

The excellent performance of the present laser to detect weak signals was demonstrated 
by a simple HHG experiment with ZnSe polycrystals. 
Figure \ref{fig5}(a) shows a typical harmonic spectra driven by 3.8 $\mu$m and 3.5 $\mu$m pulses 
generated in a 3-mm-thick ZnSe polycrystal.
One can clearly see 
harmonics up to 8th order of 3.8 $\mu$m wave below the bandgap [2.67 eV (464 nm)] 
as well as novel two peaks at $\approx 250$ nm 
and $\approx 400$ nm above the bandgap [Fig.~\ref{fig5}(b)],
which cannot be explained as simple harmonic peaks;
neither their separation nor the position were integer multiples of the photon energy of 3.8 $\mu$m pulses. 
Here by using 3.5 $\mu$m pulses, we also observed a wavelength-independent peak at 2.61 eV [red curve in Fig.~\ref{fig5}(a)]  explained 
by the luminescence from excitonic states created by inter-band multi($\ge 9$)-photon transitions. 

To understand the origins of the above two novel peaks, we calculated imaginary part of the dielectric function $\epsilon_2(\omega):=\textrm{Im} [\epsilon (\omega)]$ 
based on the nonlocal pseudopotential model \cite{PhysRevB.14.556,Adachi1991}.
The positions of the two peaks were identified to those of $\epsilon_2(\omega)$ and
their physical origins are in the so-called Van Hove singularities of the 2D exciton peak 
and one of the two 3D exciton ones.  Here the latter one is related to the 2.61 eV peak 
by the spin-orbit splitting with an energy of 0.43 eV.
The large $\epsilon_2 (\omega)$ indicates not only large absorption in the perturbative regime 
but also existence of a large number of opened quantum paths in general, it can explain the physical origin of the these peaks.
In fact, the excitonic peak below the bandgap have been observed in many materials 
as a photon emission \cite{ghimire2011observation}
and combining these signals with ultrafast spectroscopic methodology can serve as another route to observe transient band structures of solids.

In conclusion, by using a SC-seeding scheme with a 1.4 ps Yb:YAG pump laser,
 we have developed a CEP-stable KTA OPA delivering 79 fs, 3.8 $\mu$m, 4.5 W pulses
at a 100 kHz repetition rate. 
Since this duration of pump is a typical pulse duration of kW-class Yb:YAG thin disk amplifiers \cite{Nubbemeyer:17}, 
future power scaling with these kW-class amplifiers is feasible.
Total output MIR power was 17 W and this high photon flux aspect was successfully 
applied to the study of HHG in ZnSe crystals, in which faint yet novel signals above their bandgap were clearly observed.

%
%%%%%%%%%%%%%%%%%%%%%%%%%%%%%%%%%%%%%%%%%%
%%%%%%%%%%%%%%%%%%%%%%%%%%%%%%%%%%%%%%%%%%
%

\vspace{2mm}
%%%%%%%%%%%%%%%%%%%%%%%%%%%%%%%%%%%%
%\section*{Funding Information}
%%%%%%%%%%%%%%%%%%%%%%%%%%%%%%%%%%% 
\noindent
\textbf{Funding.}
The National Research Foundation 
of Korea (NRF) (No.~2018R1D1A1B07051349, No.~2009-00439, No.~2016K1A4A4A01922028).

\bibliography{optlett4}

\end{document}